\documentclass[11pt]{article}
\pdfoutput=1
\usepackage{latexsym}
\usepackage{amssymb}
\usepackage{amsmath}
\usepackage{hyperref}
\usepackage{graphicx}

\newcommand{\tr}{{\rm Tr}}

\usepackage{color}

\begin{document}
\pagestyle{empty}

\begin{flushright}
KEK-TH-1997
\end{flushright}

\vspace{3cm}

\begin{center}

{\bf\LARGE  {\boldmath$\theta = \pi$} in
 SU(\boldmath$N$)/$\mathbb{Z}_{\boldmath N}$ gauge theories}
\\

\vspace*{1.5cm}
{\large 
Ryuichiro Kitano$^{a,b}$, Takao Suyama$^{a}$ and Norikazu Yamada$^{a,b}$
} \\
\vspace*{0.5cm}

{\it 
$^a$KEK Theory Center, Tsukuba 305-0801,
Japan\\
$^b$Graduate University for Advanced Studies (Sokendai), Tsukuba
305-0801, Japan
}

\end{center}

\vspace*{1.0cm}

\begin{abstract}
{\normalsize

In ${\rm SU}(N)$ gauge theory, it is argued recently that there exists a
 ``mixed anomaly'' between the CP symmetry and the 1-form $\mathbb{Z}_N$
 symmetry at $\theta = \pi$, and the anomaly matching requires CP to be
 spontaneously broken at $\theta = \pi$ if the system is in the
 confining phase.
In this paper, we elaborate on this discussion by examining the large
 volume behavior of the partition functions of the ${\rm
 SU}(N)/\mathbb{Z}_N$ theory on $T^4$ \`{a} la 't~Hooft.  The
 periodicity of the partition function in $\theta$, which is not $2\pi$
 due to fractional instanton numbers, suggests the presence of a phase
 transition at $\theta=\pi$.  We propose lattice simulations to study
 the distribution of the instanton number in ${\rm SU}(N)/\mathbb{Z}_N$
 theories. A characteristic shape of the distribution is predicted when
 the system is in the confining phase. The measurements of the
 distribution may be useful in understanding the phase structure of the
 theory.  }
\end{abstract} 

\newpage
\baselineskip=18pt
\setcounter{page}{2}
\pagestyle{plain}
\baselineskip=18pt
\pagestyle{plain}

\setcounter{footnote}{0}

\section{Motivation}

The $\theta$-term in 4D gauge theories specifies how to sum up
topologically inequivalent sectors in the path integral. In the real
world, the value of the $\theta$ parameter in QCD is physical and known
to be unnaturally small if the up-quark mass is non-vanishing.
Unlike other parameters in the Lagrangian, the $\theta$ parameter only shows up at the
non-perturbative level. 
The reaction of the theory to the change of the
$\theta$ parameter gives us quite important information on the vacuum
structure of the theory.

The theory at $\theta = \pi$ is somewhat interesting. It is the point
where the Lagrangian has CP invariance (up to $2\pi$ shift of $\theta$),
as well as at $\theta = 0$.
A non-trivial phenomenon at $\theta = \pi$ has been found in 4D bosonic
pure Yang-Mills theory in the large $N$ limit
\cite{Witten:1980sp,Witten:1998uka}.  It is shown in
\cite{Witten:1980sp,Witten:1998uka} that the vacuum energy of the theory
depends on $\theta$ as
\begin{equation}
F(\theta) = C\min_k(\theta+2\pi k)^2+O(1/N)
\end{equation}
with a constant $C$.  At $\theta=\pi$, this function has a cusp.  This
shows that the expectation value $\langle F_{\mu\nu}\tilde{F}_{\mu\nu}
\rangle$ has a discontinuity there, indicating the spontaneous CP
violation.  One may investigate the same kind of phenomenon in a more
rigorous way for ${\cal N}=1$ super Yang-Mills theory with gauge group
${\rm SU}(N)$ when a small gaugino mass $m$ is added which breaks
supersymmetry \cite{Evans:1996hi, Konishi:1996iz, Aharony:2013hda,
Dine:2016sgq, Gaiotto:2017yup}.  The following effective potential
\begin{equation}
F(\theta) = -2m\mu^3e^{-8\pi^2/( g(\mu)^2 N ) +i\theta/N}+{\rm c.c.}, 
   \label{potential}
\end{equation}
is induced, where $\mu$ is a renormalization scale.  When
$\theta=\pm\pi$, a pair of vacua degenerate.  The potential
(\ref{potential}) implies that the expectation values $\langle
F_{\mu\nu}\tilde{F}_{\mu\nu} \rangle$ at $\theta=\pm \pi$ are nonzero
and have opposite signs with each other.  
Although the limit $m \to \infty$ is hard to study, this example
indicates that the spontaneous CP violation may occur also for finite
$N$ gauge theories.
Indeed, 't~Hooft has argued that there must be a phase transition at some
value of $\theta$ (likely to be at $\theta = \pi$) if electric
confinement takes place for all values of
$\theta$~\cite{tHooft:1981sps}. The possibility of a transition to the
oblique confinement phase near $\theta \simeq \pi$ has been proposed in
Ref.~\cite{tHooft:1981bkw}.
See Ref.~\cite{Affleck:1991tj} for the discussion on spontaneous C and
P violation at $\theta = \pi$ in 2D ${\mathbb{CP}^{N-1}}$ models.

Recently, a renewed discussion has led to the conclusion that there must
be spontaneous CP violation at $\theta = \pi$ in bosonic ${\rm SU}(N)$
Yang-Mills theory in the confining phase~\cite{Gaiotto:2017yup}.
The argument is based on the ``anomaly matching'' for the mixed anomaly
between CP symmetry and the center symmetry.  The latter is an example
of the 1-form symmetry \cite{Gaiotto:2014kfa}.  A background gauge field
for the 1-form $\mathbb{Z}_N$ symmetry breaks the $2 \pi$ periodicity
for the $\theta$ parameter, which makes it impossible to maintain the CP
invariance at $\theta = 0$ and $\theta = \pi$ simultaneously.  This
anomaly should persist at any energy scale.  In the infrared, assuming
that the theory is in the confining phase, the low energy effective
theory is trivial.  However, the trivial theory cannot produce the mixed
CP-$\mathbb{Z}_N$ anomaly, indicating that CP should be spontaneously
broken in the ${\rm SU}(N)$ gauge theory, since the 1-form
$\mathbb{Z}_N$ symmetry is preserved in the confining phase.
This anomaly matching argument has been applied to other theories in
\cite{Tanizaki:2017bam,Komargodski:2017smk,Shimizu:2017asf,Kikuchi:2017pcp} for the
discussion on their phase structure.

Motivated by the work in \cite{Gaiotto:2017yup}, we try to understand
the spontaneous CP violation in ${\rm SU}(N)$ Yang-Mills theory on
$\mathbb{R}^4$ by studying the large volume limit of ${\rm
SU}(N)/\mathbb{Z}_N$ Yang-Mills theory on $T^4$.
The discussion is along the line of the argument of
't~Hooft~\cite{tHooft:1981sps} where free energies of electric and
magnetic line operators~\cite{tHooft:1977nqb} are studied as functions
of $\theta$.
As discussed in \cite{tHooft:1979rtg, tHooft:1981nnx, vanBaal:1982ag},
there are non-trivial bundles on $T^4$ in ${\rm SU}(N)/\mathbb{Z}_N$
theory, realized by twisted boundary conditions. These topologically
inequivalent bundles should be summed up in the path integral.  Whereas
in ${\rm SU}(N)$ theory, only the trivial topology, except for
instantons, is allowed. This makes two theories, ${\rm SU}(N)$ and ${\rm
SU}(N)/\mathbb{Z}_N$ theories, different from each other.
However, in the confining phase where the correlation lengths between
local operators are finite, local physics in the two theories should
still be identical with each other when the volume of $T^4$ is large
enough.  Instead, a difference of these theories can be found in the
partition functions.
In ${\rm SU}(N)/\mathbb{Z}_N$ theory on $T^4$, magnetic line operators
are light in the confining phase since magnetic fluxes are screened.
This means that the sectors with non-trivial bundles may contribute to
the path integral even in the large volume limit.
We find that these two facts, the same local physics and different
partition functions, suggest the presence of a phase transition in ${\rm
SU}(N)$ theory on $\mathbb{R}^4$ at some value of $\theta$.

We find possible applications of our discussion to lattice gauge
theory. By putting ${\rm SU}(N)/\mathbb{Z}_N$ theory on the lattice and by
measuring the instanton numbers which can be fractional, one can extract
information on the phase structure of ${\rm SU}(N)$ theory. We
discuss what kinds of phenomena are anticipated, and also what are advantages to
study ${\rm SU}(N)/\mathbb{Z}_N$ theory rather than ${\rm SU}(N)$ theory directly.

This paper is organized as follows.  In section \ref{review}, we review
${\rm SU}(N)/\mathbb{Z}_N$ gauge theory on $T^4$, focusing on its global
structures.  The dynamical aspects of this theory are investigated in
section \ref{large volume} where we show a strong evidence of the
existence of a first order phase transition with the spontaneous CP
violation.  In section \ref{lattice}, we argue possible implications for
lattice simulation of ${\rm SU}(N)/\mathbb{Z}_N$ gauge theory on $T^4$.

\section{${\rm SU}(N)/\mathbb{Z}_N$ gauge theory on $T^4$} \label{review}

In the following, we consider a gauge theory on $T^4$ whose gauge group
is $G:={\rm SU}(N)/\mathbb{Z}_N$.  This theory is locally equivalent to
$\tilde{G}:={\rm SU}(N)$ theory.  However, the global structure of the
former theory is known to be much richer than that of the latter theory.

Let us recall the global formulation of a gauge theory on a general
manifold $M$ \cite{Nakahara:1990th}.  Let $M=\bigcup_i U_i$ be an open
covering of $M$.  For each intersection $U_{ij}:=U_i\cap U_j$, we define
a transition function $g_{ij}$ which takes its values in $G$.  The
transition functions must satisfy
\begin{equation}
g_{ij} = g_{ji}^{-1} \mbox{ on }U_{ij}, \hspace{5mm} g_{ij}g_{jk}g_{ki} = 1\mbox{ on }U_{ijk}, 
\end{equation}
where $U_{ijk}:=U_i\cap U_j\cap U_k$. 
A principal $G$-bundle is defined by gluing $U_i\times G$ using $g_{ij}$. 
The gauge field is defined as a connection on this principal $G$-bundle.

To see the difference between $G$ theory and $\tilde{G}$ theory, we
examine whether a given $G$-bundle can be regarded as a
$\tilde{G}$-bundle.  There is the canonical homomorphism
\begin{equation}
\pi\ :\ \tilde{G}\ \to\ G
\end{equation}
whose kernel is $\mathbb{Z}_N$.  For each element $g\in G$, one can
choose an element $\tilde{g}\in\tilde{G}$ such that $\pi(\tilde{g})=g$
holds.  By this procedure, and assuming the continuity, the transition
functions $g_{ij}$ can be uplifted to functions $\tilde{g}_{ij}$ which
take their values in $\tilde{G}$.  The functions $\tilde{g}_{ij}$,
however, may not define transition functions for a principal
$\tilde{G}$-bundle since the choice of $\tilde{g}\in\tilde{G}$ for a
given $g\in G$ is not unique.  In general, they satisfy
\begin{equation}
\tilde{g}_{ij}\tilde{g}_{jk}\tilde{g}_{ki} = C_{ijk} \in \mathbb{Z}_N\mbox{ on }U_{ijk}. 
\end{equation}
This is due to the fact that $\pi(C_{ijk})=1$ holds. 

There is a possibility that all $C_{ijk}$ can be set to the identity by
choosing a suitable $\tilde{g}_{ij}$ for each $g_{ij}$.  This
corresponds to multiplying a suitable $C_{ij}\in\mathbb{Z}_N$ to each
$g_{ij}$.  Then, $C_{ijk}$ changes as
\begin{equation}
C_{ijk} \to C'_{ijk}:=C_{ijk}C_{ij}^{-1}C_{jk}^{-1}C_{ki}^{-1}. 
\end{equation}
These two sets of the factors $\{C_{ijk}\}$ and $\{C'_{ijk}\}$ are
regarded as equivalent since they are obtained from the same $G$-bundle.
Therefore, each principal $G$-bundle is associated to a 2-cocycle on $M$
whose values are in $\mathbb{Z}_N$.  They are classified by the
cohomology group $H^2(M,\mathbb{Z}_N)$ \cite{Witten:2000nv}.

Consider the case $M=T^4$.  It is easy to show that
\begin{equation}
H^0(S^1,\mathbb{Z}_N) = H^1(S^1,\mathbb{Z}_N) = \mathbb{Z}_N
\end{equation}
holds. 
By K\"unneth formula, one finds 
\begin{equation}
H^2(T^4,\mathbb{Z}_N) = (\mathbb{Z}_N)^6. 
\end{equation}
This implies that there are $N^6$ kinds of distinct $G$-bundles among
which only one can be regarded as a $\tilde{G}$-bundle.

$G$-bundles on $T^4$ can be described more explicitly as
follows~\cite{tHooft:1979rtg, tHooft:1981nnx, vanBaal:1982ag}.  The
global structure of a $G$-bundle on $T^4$ comes from a twisted boundary
condition for the gauge field.  Let us define coordinates $x_\mu$ on
$T^4$ such that $x_\mu=0$ and $x_\mu=a_\mu$ are identified.  The gauge
field at $x_\mu=a_\mu$ is related to the one at $x_\mu=0$ by a gauge
transformation as
\begin{equation}
A_\lambda(x_\mu=a_\mu) = \Omega_\mu A_\lambda(x_\mu=0)\Omega_\mu^{-1}
-i\Omega_\mu\partial_\lambda\Omega_\mu^{-1}. 
\end{equation}
The compatibility conditions for $\Omega_\mu$ are 
\begin{equation}
\Omega_\mu(x_\nu=a_\nu)\Omega_\nu(x_\mu=0) = e^{2\pi in_{\mu\nu}/N}
\Omega_\nu(x_\mu=a_\mu)\Omega_\mu(x_\nu=0), 
\end{equation}
where $n_{\mu\nu}$ are integers modulo $N$. 
Note that $n_{\mu\nu}$ is anti-symmetric. 
These integers label $N^6$ distinct $G$-bundles.

The non-triviality of $G$-bundles appears in the values of the Pontryagin index. 
It is given by~\cite{tHooft:1981nnx, vanBaal:1982ag}
\begin{align}
 P& = {1 \over 16 \pi^2}\int d^4 x F_{\mu \nu} \tilde F_{\mu \nu}
= \nu + \left(
{N - 1 \over N} \right)
{n_{\mu \nu} \tilde n_{\mu \nu} \over 4},
\label{eq:index}
\end{align}
where $\nu$ is an integer and
\begin{align}
 \tilde n_{\mu \nu}& := {1 \over 2} \epsilon_{\mu \nu \rho \sigma}
 n_{\rho \sigma}.
\end{align}
For example, for $n_{12} = n_{34} = 1$ and zeros for other components,
we have 
\begin{align}
 P& = \nu + {N-1 \over N}.
   \label{fractional P}
\end{align}
Therefore, the index is not an integer in general.

The Pontryagin index appears in the action of $G$ theory as the $\theta$-term. 
The partition function $Z_G(\theta,V)$ is given as 
\begin{equation}
Z_G(\theta,V) = \sum_{n_{\mu\nu},\nu}Z_G(n,\nu,V)e^{iP\theta}
\label{eq:zG}
\end{equation}
where $Z_G(n,\nu,V)$ is the path integral over the sector with fixed
$n_{\mu\nu}$ and $\nu$, in which the action does not contain the
$\theta$-term.  Since $P$ takes fractional values (\ref{fractional P}),
$Z_G(\theta,V)$ has $2 N \pi$ periodicity.  On the other hand, the
partition function of $\tilde{G}$ theory is
\begin{equation}
Z_{\tilde{G}}(\theta,V) = \sum_{\nu}Z_G(0,\nu,V)e^{i\nu\theta}
\label{eq:zGtilde}
\end{equation}
which has $2\pi$ periodicity.

The integers $n_{\mu\nu}$ have the following physical meaning. 
Suppose that $\mu=4$ corresponds to the time direction. 
Then $n_{\mu\nu}$ can be decomposed into 
\begin{equation}
k_i := n_{i4} \hspace{5mm} \mbox{and} \hspace{5mm} m_i := \frac12\varepsilon_{ijk}n_{jk}, 
\end{equation}
where $i,j,k=1,2,3$. 
In the following, we denote the path integral with fixed $k_i,m_i,\nu$ as $Z_G(k,m,\nu,V)$. 

The 3-vector $m_i$ is called the magnetic flux.  This name is justified
by observing that $m_i$ changes by one if an 't~Hooft line operator
along the $i$-th direction is inserted.  Then, one might expect that
$k_i$ would be interpreted as the electric flux.  However, this turns
out not to be the case.  Instead, another 3-vector $e_i$ which appears
in the following expression \cite{tHooft:1981sps}
\begin{equation}
e^{-VF_G(e,m,\theta,V)}=\frac1{N^3}\sum_{k,\nu}e^{-2\pi ik_ie_i/N+i\theta(\nu-k_im_i/N)}Z_G(k,m,\nu,V),
   \label{Fourier}
\end{equation}
is called the electric flux.  This is because $e_i$ changes by one if a
Wilson line operator along the $i$-th direction is inserted.  The
quantity $F_G(e,m,\theta,V)$ in the left-hand side is the free energy
density for a sector with fixed $e_i$ and $m_i$.

In the canonical formalism, the fluxes appear as follows
\cite{Witten:2000nv}.  The Hilbert space of $G$ theory on $T^4$ is the
space of wave functions on the configuration space which is the space of
connections on a $G$-bundle on the spatial manifold $T^3$.  The
$G$-bundles on $T^3$ are classified by
$H^2(T^3,\mathbb{Z}_N)=(\mathbb{Z}_N)^3$.  They are labeled by the
magnetic flux $m_i$.  In defining the partition function on $T^4$, one
may insert a twist in the time-direction.  The twist is a gauge
transformation which is specified by $e_i$.

\section{Large volume limit}   \label{large volume}

Local physical quantities, such as $n$-point functions of local
operators, should become independent of the volume $V$ of $T^4$ when $V$
is large enough.  More precisely, if the theory under consideration has
a mass gap $\Delta>0$, then the volume-independence is expected when the
size $V^{1/4}$ of $T^4$ is much larger than $\Delta^{-1}$.  Therefore,
assuming the existence of a mass gap, we expect that all the local
quantities in $G$ theory on $T^4$ in the large volume limit should
coincide with those in $\tilde{G}$ theory on $\mathbb{R}^4$.

The spontaneous CP violation in $\tilde{G}$ theory on $\mathbb{R}^4$ can
be probed by the expectation value $\langle F_{\mu\nu}\tilde{F}_{\mu\nu}
\rangle_{\tilde{G}}$.  In the following, instead, we investigate the
$\theta$-dependence of $\langle F_{\mu\nu}\tilde{F}_{\mu\nu}
\rangle_{G,V}$ in $G$ theory on $T^4$.  In the presence of a mass gap,
the finite volume correction is exponentially suppressed
\cite{Luscher:1985dn, Bijnens:2006ve}:
\begin{equation}
\langle F_{\mu\nu}\tilde{F}_{\mu\nu} \rangle_{G,V}-\langle F_{\mu\nu}\tilde{F}_{\mu\nu} \rangle_{G,\infty} = O\left( \exp(- \Delta V^{1/4}) \right). 
   \label{finite volume effect}
\end{equation}
This can be understood as follows.  A theory on $T^4$ is regarded as the
same theory on $\mathbb{R}^4$ with mirror images.  The finite volume
correction then comes from interactions with the mirror images which are
suppressed exponentially as (\ref{finite volume effect}) since the
distance to the nearest image is of order $V^{1/4}$.  Since the
difference between $G$ theory and $\tilde{G}$ theory comes from the
global structure discussed in section \ref{review}, the quantity
$\langle F_{\mu\nu}\tilde{F}_{\mu\nu} \rangle_{G,\infty}$ should
coincide with $\langle F_{\mu\nu}\tilde{F}_{\mu\nu}
\rangle_{\tilde{G}}$.  Therefore, if a phase transition would exist in
$G$ theory, then this implies the existence of the same phase
transition in $\tilde G$ theory which results in the spontaneous CP violation.

\subsection{The partition function}

Recall that the partition function $Z_{G}(\theta,V)$ of $G$ theory is a
periodic function of $\theta$ with period $2N\pi$ when $V$ is finite.
One might naively expect that the period of $Z_{G}(\theta,V)$ would
become $2\pi$ in the large volume limit since the effect of the twisted
bundles would become irrelevant in the limit.

To clarify this issue, let us consider $Z_G(\theta,V)$ in more detail.
The relation (\ref{Fourier}) implies
\begin{equation}
\sum_{\nu}e^{i\theta(\nu-k_im_i/N)}Z_G(k,m,\nu,V)\ =\ \sum_{e}e^{2\pi ik_ie_i/N}e^{-VF_G(e,m,\theta,V)}. 
\end{equation}
Using this relation and Eq.~\eqref{eq:zG}, $Z_G(\theta,V)$ can be
written as
\begin{equation}
Z_G(\theta,V) = N^3\sum_{m}e^{-VF_G(0,m,\theta,V)}, 
\end{equation}
where the summations over $k$ and $e$ have been performed. 

We are interested in the values $Z_G(2l\pi,V)$ with $l=0,1,\cdots,N-1$. 
The relation (\ref{Fourier}) implies 
\begin{equation}
F_G(e,m,\theta+2l\pi,V) = F_G(e+lm,m,\theta,V), 
\end{equation}
which is nothing but the Witten effect \cite{Witten:1979ey}. 
This relation then implies 
\begin{equation}
Z_G(2l\pi,V) = N^3\sum_{m}e^{-VF_G(lm,m,0,V)}. 
\end{equation}
Note that the fluxes appearing in the sum correspond to the line
operators which are allowed to exist in the
theories~\cite{Aharony:2013hda, Kapustin:2005py, Gaiotto:2010be}.  This
is also the case for the partition function of $\tilde{G}$ theory in
Eq.~\eqref{eq:zGtilde}:
\begin{equation}
Z_{\tilde{G}}(\theta,V) = \sum_ee^{-VF_G(e,0,\theta,V)}. 
\end{equation}

In the case $l=0$, the fluxes in the sum are purely magnetic.  In the
confining phase, they are light, meaning that
$F_G(0,m,0,V)-F_G(0,0,0,V)$ vanish exponentially in the large volume
limit~\cite{tHooft:1979rtg}.  Therefore, $Z_G(0,V)$ becomes
\begin{equation}
Z_G(0,V) \sim N^6e^{-VF_G(0,0,0,V)}
\end{equation}
for a large enough volume.  On the other hand, the fluxes in the cases
$l\ne0$ are heavy, {\it i.e.,} the free energies do not vanish
exponentially, except for $m=0$.  Since the contributions from the heavy
fluxes are negligible, we obtain
\begin{equation}
Z_G(2\pi l,V) \sim N^3e^{-VF_G(0,0,0,V)}. \hspace{5mm} (l=1,2,\cdots,N-1)
\end{equation}
This result clearly indicates that $Z_G(\theta,V)$ has $2N\pi$
periodicity even in the large volume limit.  The ratio
\begin{equation}
Z_G(2\pi l,V)/Z_G(0,V) \sim N^{-3}
\end{equation}
gives the ratio of the numbers of light fluxes.

The full partition function is also given by a Fourier series as
follows:
\begin{align}
Z_{G} (\theta, V)& = \sum_{P} c_P(V) e^{i P \theta}, \quad P =
 0,\  \pm {1 \over N},\  \pm{2 \over N}, \ \cdots.
 \label{eq:partitionf}
\end{align}
where the coefficients $c_P(V)$ are given as 
\begin{equation}
c_P(V) = \sum_{\nu+\frac{N-1}{N}k_im_i=P}Z_G(k,m,\nu,V). 
\end{equation}
They are partition functions of the sector with a fixed $P$ for $\theta
= 0$, and thus they are real and positive (in the Euclidean theory).
The existence of non-vanishing coefficients $c_P (V)$ with fractional
indices $P$ is expected from the $2N\pi$ periodicity of $Z_G(\theta,V)$
even in the infinite volume limit, as long as the theory is in the
confining phase at $\theta = 0$.

\subsection{Mass gap} \label{mass gap}

The results obtained so far can be applied to the free energy density,
which we write as
\begin{equation}
F_{G}(\theta,V):=-\frac1V\log Z_{G} (\theta, V) = F_{G}(\theta,\infty)+g(\theta,V), 
   \label{free energy density}
\end{equation}
where $g(\theta,V)$ represents the finite size correction.  It was found
that $F_G(\theta,V)$ is $2N\pi$ periodic, while $F_G(\theta,\infty)$
should be $2\pi$ periodic as it should coincide with the free energy in
$\tilde G$ theory up to a constant.  Therefore, the finite size
correction $g(\theta,V)$ has $2N\pi$ periodicity.  Quantitatively, we
found in the previous subsection that
\begin{equation}
g(2\pi,V)-g(0,V) \sim \frac{\log N^3}V
   \label{gap}
\end{equation}
holds for a large enough volume. 

The $\theta$-derivative of $F_G(\theta,V)$ gives us the expectation value 
\begin{align}
{1 \over 16 \pi^2}\langle F_{\mu \nu} \tilde F_{\mu \nu} \rangle_{G,V}
& =
 - {i \over V} {\partial \over \partial \theta} \log Z_{G}
 (\theta, V)
 = i \frac{\partial}{\partial\theta}F_{G}(\theta,\infty) + {i {\partial \over \partial \theta} g(\theta, V) }. 
\end{align}
Since this is an expectation value of a local operator, its finite size
correction $i\partial_\theta g(\theta,V)$ should depend on the volume as
$\exp(- \Delta V^{1/4})$~\cite{Luscher:1985dn, Bijnens:2006ve} in the
presence of a mass gap $\Delta$. (See Eq.~\eqref{finite volume effect}.)
Therefore, the derivative $\partial_\theta g(\theta, V)$ should be
exponentially suppressed at large $V$, implying that $g(\theta,V)$
should be almost a constant in $\theta$.

The above arguments have the following consequence.  Since $g(\theta,V)$
is an almost constant function satisfying (\ref{gap}), a natural
expectation for the functional form would be
\begin{equation}
g(\theta,V) \sim \left\{
\begin{array}{cc}
g(0,V), & (0\le\theta\lesssim \theta_c) \\ [2mm]
g(0,V)+\displaystyle{\frac{\log N^3}{V}}, & (\theta_c\lesssim\theta\le 2\pi)
\end{array}
\right.
   \label{jump}
\end{equation}
for large enough $V$.  By symmetry, we expect $\theta_c=\pi$.  If this
is indeed the case, then this indicates that the free energy density
$F_G(\theta,V)$ increases abruptly around $\theta=\pi$, indicating that
the finite size correction $i \partial_\theta g(\theta,V)$ becomes much larger than
expected from the one due to the mass gap.

\subsection{Phase transition}

The appearance of a large finite size effect is a typical signature of a
phase transition as we discuss below.
In general, when we have a first-order phase transition, the partition
function near the critical point $\theta_c$ has the following form:
\begin{align}
Z (\theta) \sim a_1 e^{-V f_1(\theta)} + a_2 e^{-V f_2(\theta)},
   \label{1st order}
\end{align}
where $a_{1,2}$ are some coefficients of $O(1)$ and $f_{1,2}$ are free
energy densities of two phases which satisfy
\begin{equation}
f_1(\theta_c) = f_2 (\theta_c), \hspace{5mm} f_1'(\theta_c)\ne f'_2(\theta_c). 
   \label{1st order condition}
\end{equation}
Here, the critical point $\theta_c$ is defined as the one in the large
$V$ limit, and possible $\theta$ and $V$ dependencies of $a_{1,2}$ can
be ignored in the following discussion at a large enough volume.

Let us consider the log-derivative: 
\begin{equation}
F'(\theta) := -\frac1V\frac{\partial}{\partial \theta}\log Z(\theta)
\sim {
a_1 f_1'(\theta) e^{-V f_1(\theta)} + a_2 f_2'(\theta) e^{-V f_2(\theta)}
\over
a_1 e^{-V f_1(\theta)} + a_2 e^{-V f_2(\theta)}
}
,
\end{equation}
which corresponds to the expectation value $\langle F_{\mu\nu} \tilde
F_{\mu\nu} \rangle_{G,V}$.  For large $V$, $F'(\theta)$ is equal to
$f_1'(\theta)$ or $f_2'(\theta)$ up to exponentially suppressed terms if
$\theta$ is away from $\theta_c$.  
For $\theta \sim \theta_c$, on the other hand, $F'(\theta)$ is
approximately given by
\begin{align}
F'(\theta) \sim
{
a_1 f_1'(\theta) e^{-V f_1' (\theta_c) (\theta - \theta_c)}
+ a_2 f_2'(\theta) e^{-V f_2' (\theta_c) (\theta - \theta_c)}
\over
a_1 e^{-V f_1' (\theta_c) (\theta - \theta_c)}
+ a_2 e^{-V f_2' (\theta_c) (\theta - \theta_c)}
}
.
\end{align}
One can see that, in a region where $|\theta -
\theta_c| < O(1/V)$ in the unit of a typical energy scale $\Delta$, the
value of $F'(\theta)$ can deviate by $O(1)$ from both $f_1'(\theta)$ and
$f_2'(\theta)$.

Applying this argument to the partition function $Z_G(\theta,V)$, it is
concluded that $\partial_\theta g(\theta,V)$ can become $O(1)$ quantity
in the region $|\theta-\pi|<O(1/V)$.  Then, integration of
$\partial_\theta g(\theta,V)$ reproduces the expected functional form
of $g(\theta,V)$ in Eq.~(\ref{jump}).  This strongly suggests that a first
order phase transition exists at $\theta=\pi$ in $G$ theory.

In the infinite volume limit, the region where $g(\theta,V)$ can vary
disappears, and $\langle F_{\mu\nu} \tilde F_{\mu\nu} \rangle_{G,V}$
develops a discontinuity at $\theta=\pi$ due to the condition (\ref{1st
order condition}).
Since $\langle F_{\mu\nu} \tilde F_{\mu\nu} \rangle_{G,\infty}$
coincides with $\langle F_{\mu\nu} \tilde F_{\mu\nu}
\rangle_{\tilde{G}}$, we conclude that the spontaneous CP violation
occurs in $\tilde{G}$ theory on $\mathbb{R}^4$ at $\theta=\pi$ since
$\langle F_{\mu\nu} \tilde F_{\mu\nu} \rangle_{\tilde{G}}$ is
discontinuous there, as in the example of the softly broken ${\cal N}=1$
super Yang-Mills theory as well as large $N$ theory.
Remember that $g(\theta, V)$ is a finite volume effect which disappears
in the large volume limit.
Nevertheless, it is interesting to note that its $\theta$ and $V$
dependencies tell us that a quantity on ${\mathbb R}^4$, $\langle
F_{\mu\nu} \tilde F_{\mu\nu} \rangle_{\tilde{G}}$, need to have a
certain property; compatibility between the periodicity of the partition
function encoded in $g(\theta, V)$ and the mass gap requires a
discontinuity in $F'(\theta, \infty)$.

The conclusion we obtained is consistent with the discussion based on
the anomaly matching~\cite{Gaiotto:2017yup}, there the cases of even and
odd $N$ are separately discussed. See also Ref.~\cite{Kikuchi:2017pcp}
for detailed discussion on the case with odd $N$.
Our discussion confirms that the same conclusion, spontaneous CP
violation at $\theta = \pi$, can be derived independent of $N$ under
the assumptions that the mass gap persists for all values of $\theta$
and a phase transition happens only once between $\theta = 0$ and $\theta = 2
\pi$.

We stress that the change (\ref{gap}) of $g(\theta,V)$ as we vary
$\theta$ is a physical observable.  First of all, {\it it can be
measured by a lattice simulation}.  This will be discussed in the next
section.  In addition, it has the following physical meaning.  Recall
that $N^3$ is the number of light fluxes at $\theta=0$.  In the large
volume limit, the light fluxes have almost zero energy, so $N^3$ can be
regarded as the partition function of a statistical system of the light
fluxes.  Therefore, the change in $g(\theta,V)$ corresponds to the
change in the entropy of these fluxes.

\subsection{Other possibilities}

So far, we have assumed the existence of a mass gap for any value of
$\theta$.  There is also a possibility, mainly for $\tilde{G}={\rm
SU}(2)$, that the mass gap disappears, {\it i.e.,} the deconfinement
transition happens, at some point $\theta = \theta_c$ or in some region.

Suppose that $\Delta$ varies with $\theta$ as 
\begin{equation}
\Delta = O(|\theta-\theta_c|^a), \hspace{5mm} a>0, 
\end{equation}
in the vicinity of the critical point $\theta=\theta_c$.  As usual, we
expect that the finite size effect would be relevant when
$1/\Delta>V^{1/4}$ is satisfied.  Then, the range of $\theta$ in which
the finite size effect can become large is
\begin{equation}
|\theta-\theta_c| < O(V^{-1/4a}). 
   \label{range-gapless}
\end{equation}
As was found in subsection \ref{mass gap}, $g(\theta,V)$ must change by
the amount $O(1/V)$ within this range.  Then, the estimate of
$\partial_\theta g(\theta,V)$ around $\theta=\theta_c$ is
$O(V^{1/(4a)-1})$.  In the case $a>1/4$, the contribution from
$g(\theta,V)$ to $\langle F_{\mu\nu} \tilde F_{\mu\nu} \rangle_{G,V}$
becomes negligible in the large $V$ limit.  Therefore, $\langle
F_{\mu\nu} \tilde F_{\mu\nu} \rangle_{G}$ can be continuous in
$\theta$, even though $g(\theta,V)$ jumps at $\theta=\theta_c$.  That
is, it would be possible that the deconfinement transition happens at
$\theta = \pi$ without spontaneous CP violation.

If $\Delta$ vanishes in a finite range of $\theta$, the phase transition
happens twice at $\theta = \theta_c$, $(0 < \theta_c < \pi)$, and
$\theta = 2 \pi - \theta_c$.
In this case, our argument so far cannot apply in the range $\theta_c <
\theta < 2 \pi - \theta_c$, and, for example, $\langle F_{\mu\nu} \tilde
F_{\mu\nu} \rangle_{\tilde{G}} = 0$ at $\theta = \pi$ would be possible.

In summary, for consistency between the estimate of the finite size
corrections and the confinement at $\theta = 0$, one either needs a
first order phase transition or the disappearance of a mass gap at some
value of $\theta_c$, $0 < \theta_c \le \pi$. In the case of the first
order phase transition at $\theta_c = \pi$, there is spontaneous CP
violation at $\theta = \pi$. In other cases, the CP invariant vacuum at
$\theta = \pi$ is possible.

\section{Lattice study of ${\rm SU}(N)/\mathbb{Z}_N$ theories}
\label{lattice}

There have been efforts to investigate large $\theta$ behavior of ${\rm
SU}(N)$ theories at zero and finite temperatures on the
lattice~\cite{Bhanot:1983sn, DElia:2012pvq, DElia:2013uaf}. However,
directly studying the $\theta = \pi$ point is practically quite
difficult due to the sign problem. The complex phase, $e^{i \theta P}$,
prevents us from interpreting the integrand of the path integral as
probabilities in Monte Carlo simulations. Instead, one can perform the
path integral at $\theta = 0$ and take the summation over the instanton
numbers later with the phase factor as a weight to obtain a path
integral at non-zero $\theta$ as in Eq.~\eqref{eq:partitionf}. But, at
$\theta = \pi$, the sum involves a numerical cancellation of order
$e^{-V \Delta^4 \pi^2 }$, which makes it numerically and statistically
not possible to obtain a meaningful result.
The currently available techniques to simulate finite $\theta$ are based
on analytic continuation from imaginary values of
$\theta$~\cite{DElia:2012pvq, DElia:2013uaf, Azcoiti:2002vk,
Panagopoulos:2011rb} or the reweighting method we just
mentioned~\cite{Bhanot:1983sn, DElia:2013uaf}.
Both methods work in a limited region close to $\theta = 0$.

We propose below a lattice study of the $\theta$ dependence of $G$
theory rather than $\tilde G$ theory. Although the sign problem is as
severe as $\tilde G$ theory, the knowledge obtained in the previous
section makes characteristic predictions on the distribution of indices
$P$ at $\theta = 0$. Using the knowledge as inputs, one should be able
to improve the statistical uncertainties compared to the study of
$\tilde G$ theory. Also, by measuring the $P$ distributions, one should
be able to exclude the possibility of the phase transition significantly
below $\theta = \pi$ and also the case of the disappearing mass gap with
a very large critical exponent.

On the lattice, $\tilde G$ and $G$ theories are formulated differently;
the link variables are constructed as the fundamental and the adjoint
representations of $\tilde G$, respectively~\cite{Halliday:1981te,
Creutz:1982ga}.
The partition function of the $G$ theory is expressed as a path
integral over the link variable in the adjoint representation, $U_A$, as
\begin{align}
 Z_G (\theta, \beta)& = \int {\cal D} U_A e^{- \beta S[U_A] + i \theta P[U_A]},
\end{align}
where
\begin{align}
 S[U_A]& = \sum_{\rm plaquette} 
\left( 1 -  {1 \over N^2 -1} \tr \ U_A^{\rm P}
\right),
\end{align}
and $P[U_A]$ is the Pontryagin index calculated based on $U_A$ which we
discuss later. The matrix $U_A^{\rm P}$ is the plaquette action made of
$U_A$. The partition function, $Z_G (\theta, V)$, is obtained by taking
the continuum limit, $\beta \to \infty$, while the space-time volume,
$V$, fixed.

In the actual simulation, one can use the link variable in the
fundamental representation by using a relation between the characters in
the adjoint and the fundamental representation; the trace of a group
element, $g$, in the adjoint representation, $\tr D_A(g)$, can be
expressed as $| \tr D_F(g) |^2 -1$ where $D_F(g)$ is the same element in
the fundamental representation~\cite{Halliday:1981te, Creutz:1982ga}.
From this relation, the action of the $G$ theory is given
by the link variable $U_F$ in the fundamental representation as
\begin{align}
 S[U_A] 
& = \sum_{\rm plaquette} \left(
{N^2 \over N^2 - 1} - {1 \over N^2 - 1} | \tr  U_F^{\rm P} |^2
\right),
\end{align}
where $U_F^{\rm P}$ is the plaquette action made of $U_F$. The path integral
measures are the same for the adjoint and the fundamental
representations.
In this formulation, we do not expect an increase of the computational
cost compared to the $\tilde G$ theory since the size of
the matrix to be integrated remains the same.

As we discussed above, simulations with a large finite $\theta$ are not
practically easy.
Instead, by the simulation at $\theta = 0$, one can measure the
partition function of each $P$ sector, $c_P(V)$ in \eqref{eq:partitionf},
as the probability of obtaining configurations with the index $P$.
The index can be measured in each configurations by counting zero modes
of the Dirac operator in the adjoint representation through the index
theorem,
\begin{align}
 P = {n_+ - n_- \over 2 N},
\end{align}
where $n_\pm$ are the number of zero modes with $\pm$ chiralities, and
they should be even numbers as each mode is accompanied by its charge
conjugate pair. The definition provides us with the index with the unit
$1/N$. 
On the lattice, the Dirac operator which maintains the index theorem has
been explicitly constructed~\cite{Neuberger:1998wv}, and by using the
definition, the appearance of the configurations with fractional
indices has been confirmed in ${\rm SU}(2)$ theory when the lattice
spacing is finite~\cite{Edwards:1998dj}, while it disappears in the
continuum limit as expected~\cite{Fodor:2009nh}.
As we discussed before, such configurations should remain unsuppressed
even in the continuum limit in the ${\rm SU}(2)/\mathbb{Z}_2 \simeq {\rm
SO}(3)$ theory
in the confining phase.

The distribution of $P$, $c_P(V)$, has information of phase structures
and the $\theta$ dependence of $G$ and $\tilde G$ theories. It is
expected that the configurations with fractional $P$'s frequently appear
in the confining phase, but do not show up in the deconfining
phase. Therefore, by heating up the system, by controlling the length of
the temporal direction, we expect to see the suppression of the
fractional $P$ configurations at the deconfining temperature.

In the confining phase, interesting numbers to calculate are
\begin{align}
 {Z_G (2 l \pi, V) \over Z_G (0,V)} = {
{\displaystyle 
\sum_P c_P(V) e^{2 \pi i l P}}
\over
{\displaystyle \sum_P c_P (V) }
},\quad l = 1, \cdots, N-1,
\end{align}
which are given by $1/N^3$ for all $l$ in the large volume limit.  The
finite size correction to $1/N^3$ is suppressed exponentially.
This prediction does not depend on the detail of the phase structure
along the $\theta$ direction as long as the theory is in the confining
phase at $\theta = 0$. 
It is an interesting observable which characterizes the confinement.

\begin{figure}[t]
 \begin{center}
  \includegraphics[width=15cm]{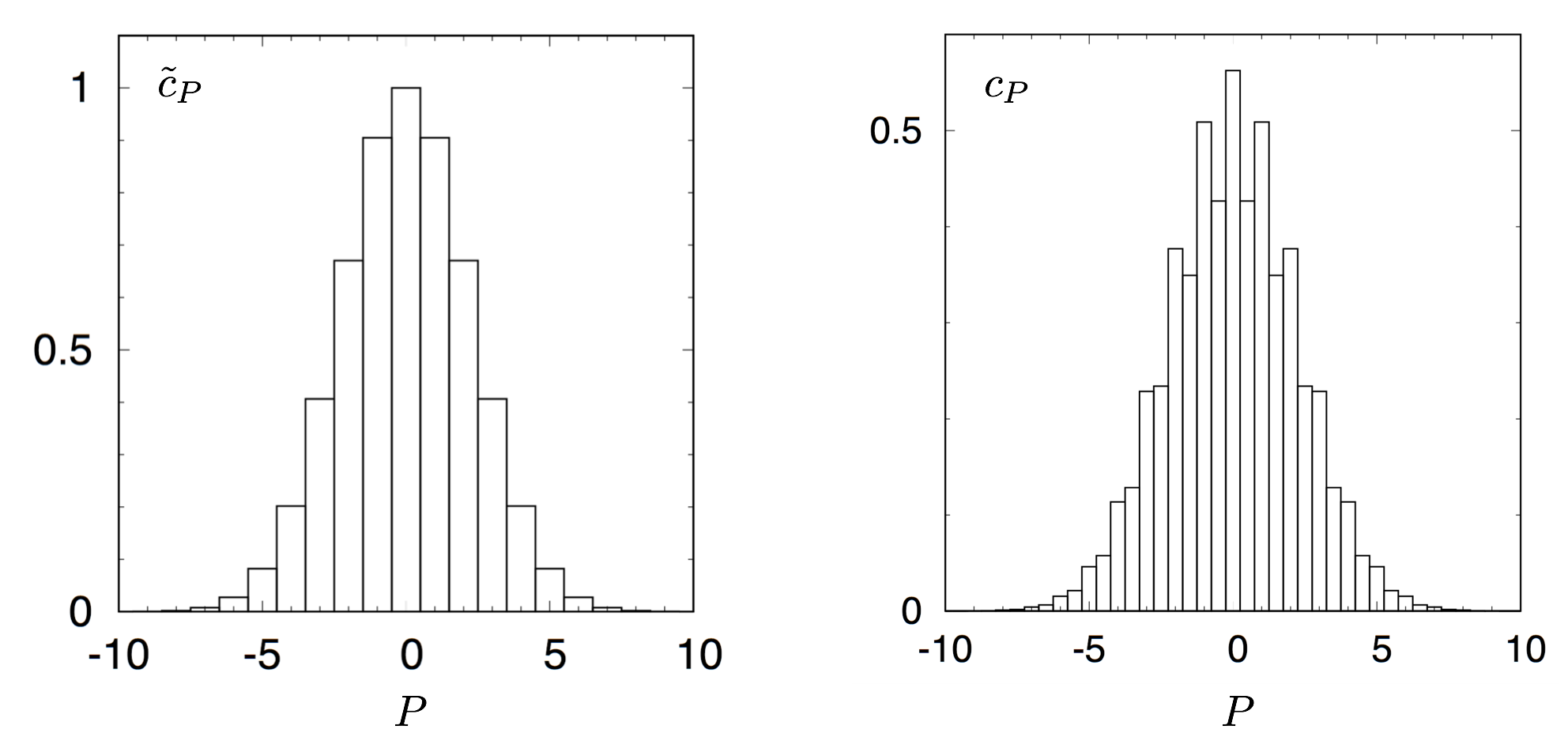}
 \end{center}
\caption{An example of the relation between $\tilde c_P$ (left) and
 $c_P$ (right) for $N=2$.}
\label{fig:cP}
\end{figure}

Let us discuss the $P$ distribution at $\theta = 0$, $c_P(V)$, in more
detail. As we discussed in the previous section, the partition functions
of the $G$ and $\tilde G$ theories at a sufficiently large volume are
related by
\begin{align}
 Z_G (\theta, V) & \sim  
\left \{
\begin{array}{ll}
 N^6 Z_{\tilde G} (\theta, V), & \quad |\theta| \lesssim \pi, \\
 N^3  Z_{\tilde G} (\theta, V), & \quad \pi \lesssim |\theta| \le N \pi. \\
\end{array}
\right. 
\label{eq:Gtheta}
\end{align}
where
\begin{align}
 Z_{\tilde G} (\theta, V) = \sum_P \tilde c_P (V) e^{i \theta P},\quad
P = 0,\ \pm 1,\ \pm 2,\ \cdots.
\end{align}
From this relation, one can express $c_P(V)$ in terms of $\tilde c_{P}
(V)$ as follows:
\begin{align}
 c_{P} (V)& \propto \left \{
\begin{array}{ll}
 {\displaystyle
{1 \over N}
\left(
1 + {1 \over N^2} - {1 \over N^3}
\right) \tilde c_P(V)}, &   \quad P : {\mbox{integer}}, \\
 {\displaystyle {1 \over N}
\left(
1 - {1 \over N^3}
\right)
\sum_{P'} \tilde c_{P'}(V) 
{
\sin \left[ \pi (P - P') \right]
\over
\pi (P - P')
}}, & \quad P : {\mbox{non-integer}}.
 \end{array}
\right.
\label{eq:cP}
\end{align}
The overall normalization is not important.
By using this formula, once we measure $c_P(V)$ with integer valued $P$,
one can predict $c_P(V)$ for non-integer $P$. 
The prediction is again not very sensitive to the behavior of
$Z_{G}(\theta, V)$ around $\theta = \pi$ since the value there is anyway
suppressed exponentially by the volume.
A significant deviation from the above prediction means that
$Z_G(\theta, V)$ is deviated from Eq.~\eqref{eq:Gtheta} in a wide range
of $\theta$, that indicates either a phase transition at a small value
of $\theta$ or a very large critical exponent.
We show in Fig.~\ref{fig:cP} the distribution predicted in
Eq.~\eqref{eq:cP} for $N=2$ by assuming that $\tilde c_P(V)$ is
gaussian.
We see an interesting non-smooth structure for small $|P|$. 

Since there are $N$ times more data points compared to $\tilde c_P(V)$,
working in $G$ theory should be advantageous in studying the $P$
distribution, {\it i.e.,} the $\theta$ dependence of the theory.
For example, when we parameterize the free energy in $\tilde G$ theory
as
\begin{align}
 F_{\tilde G} (\theta, V \to \infty) = {\chi_t \over 2} \theta^2
\left(
1 + b_2 \theta^2 + b_4 \theta^4 + \cdots
\right),
\end{align}
the parameters, $\chi_t$, $b_2$, $b_4$, $\cdots$ can be determined by
fitting the $c_P(V)$ distribution. Compared to fitting with the $\tilde
c_P(V)$ distribution, one can use the constraints in Eq.~\eqref{eq:cP},
which should help us to reduce the statistical uncertainties.

\section*{Acknowledgements}

We would like to thank Ryosuke Sato, Hideo Matsufuru and Michael Dine
for discussions and Nati Seiberg for reading the manuscript.
This work is supported by JSPS KAKENHI Grant No.~15H03669 (RK, NY),
15KK0176 (RK) and 16H06490 (TS), MEXT KAKENHI Grant No.~25105011 (RK),
the Large Scale Simulation Program No.~16/17-28 of High Energy
Accelerator Research Organization (KEK), and Interdisciplinary
Computational Science Program No.~17a15 in CCS, University of Tsukuba.

\end{document}